\documentclass[floatfix,aps,preprintnumbers,amsmath,amssymb,nofootinbib,superscriptaddress]{revtex4}
\usepackage[T1]{fontenc}
\usepackage{lmodern}
\usepackage{graphicx}
\usepackage{amsmath,amssymb,bm}
\usepackage{physics}
\usepackage{xcolor}
\usepackage{comment}
\usepackage{multirow}
\usepackage{hyperref}
\hypersetup{colorlinks=true, linkcolor=blue, citecolor=blue, urlcolor=blue}

\newcommand{\bea}{\begin{eqnarray}}
\newcommand{\eea}{\end{eqnarray}}
\newcommand{\be}{\begin{equation}}
\newcommand{\ee}{\end{equation}}
\newcommand{\np}{{\bf p}}

\newcommand{\nh}{{\bf h}}

\begin{document}

\title{Microscopic calculation of two-particle-two-hole meson-exchange currents in $^{40}$Ar and asymmetric scaling properties for neutrino and electron scattering}

\author{V.L. Martinez-Consentino}\email[]{victormc@ugr.es}
\affiliation{Departmento de Ciencias Integradas, Universidad de Huelva, E-21071 Huelva, Spain.}
\affiliation{Departamento de Sistemas F\'isicos, Qu\'imicos y Naturales, Universidad Pablo de Olavide, E-41013 Sevilla, Spain}

\author{J. Segovia}
\email[]{jsegovia@upo.es}
\affiliation{Departamento de Sistemas F\'isicos, Qu\'imicos y Naturales, Universidad Pablo de Olavide, E-41013 Sevilla, Spain}

\author{J.E. Amaro}\email[]{amaro@ugr.es} 
\affiliation{Departamento de F\'{\i}sica At\'omica, Molecular y Nuclear, Universidad de Granada, E-18071 Granada, Spain.}
\affiliation{Instituto Carlos I de F{\'\i}sica Te\'orica y Computacional, Universidad de Granada, E-18071 Granada, Spain.}

\date{\today}

\begin{abstract}
We present a microscopic calculation of two particle-two hole meson exchange current response functions in asymmetric nuclei, with particular emphasis on the $^{40}$Ar nucleus. Employing a relativistic mean-field and relativistic Fermi gas framework, we compute the nuclear response for $^{40}$Ar and compare it with that of the symmetric $^{40}$Ca nucleus, analyzing the role of proton-neutron imbalance. The model incorporates distinct proton and neutron Fermi momenta to accurately capture the nuclear dynamics of systems with $Z \neq N$. Our results indicate that using $^{40}$Ca as a proxy for $^{40}$Ar leads to a systematic error of approximately 10\%. Additionally, we propose an asymmetric scaling formula to obtain the 2p2h response for arbitrary nuclei from the $^{12}$C response, improving the description of asymmetric nuclei. Finally, we benchmark our predictions against inclusive electron scattering and neutrino cross sections.
\end{abstract}

\maketitle

\section{Introduction}
Recent neutrino experiments highlight the need to improve the precision of neutrino interaction models to enable more accurate interpretation of oscillation measurements and to reduce systematic uncertainties~\cite{Mos16,Rus18}. New experiments, such as DUNE~\cite{Dune16} and MicroBooNE~\cite{Micro25}, employ argon targets to increase statistics and enhance control over the observables of the final-state particles. However, nuclear modeling efforts have traditionally focused on symmetric nuclei such as $^{12}$C~\cite{Nie04,Mar09,Roc15}, and the use of asymmetric systems, where $Z \ne N$, introduces additional complexity~\cite{Bar18}. As a result, achieving reliable predictions of the cross section for a wide range of nuclei -especially asymmetric ones- continues to pose an outstanding challenge.

A major difficulty in modeling the charged-current (CC) neutrino-nucleus cross section across the energy range from a few hundred MeV up to several GeV is the need to account for multiple reaction channels: quasielastic (QE) single-nucleon knockout, multinucleon emission (such as 2p2h and 3p3h), pion production ($1\pi$), and deep inelastic scattering (DIS). Among these, 2p2h processes play a particularly crucial role. Recent measurements of two-proton final states in MicroBooNE~\cite{Micro25} have shown that an accurate description of 2p2h contributions in $^{40}$Ar is essential for reliable neutrino energy reconstruction.

In addition to the importance of a reliable treatment of 2p2h mechanisms in asymmetric nuclei such as $^{40}$Ar, we note that recent studies have also examined the role of interference between one-body and two-body currents in the quasielastic channel, see for example Ref. \cite{Lov25}. These interference terms are not commonly included in most existing models, and their quantitative impact is the subject of ongoing research. A detailed investigation of such effects lies beyond the scope of the present work, which focuses on the explicit calculation of the 2p2h response in asymmetric systems.

Significant theoretical uncertainties persist in the modeling of 2p2h contributions. Several microscopic approaches have been developed, including those by Nieves \textit{et al.}~\cite{Nie11}, the SuSAv2+MEC model~\cite{Meg16}, the Ghent model~\cite{Van17}, the GiBUU approach~\cite{Mos14}, the Lyon model~\cite{Rus25},the Extended Factorization approach \cite{Roc18}, and the more recent RMF model~\cite{Mar21a}. Although these models partially describe experimental data---due to the experimental difficulty of measuring the 2p2h channel---they differ significantly in their underlying assumptions and predictions. A clear example of this is the different treatment of the $\Delta(1232)$ resonance in each model~\cite{Ose87,Dek94,Kim96}, or the choice of alternative values for the nucleon to $\Delta$ axial transition form factors~\cite{Her07, Che23}.

Some of these models have been incorporated into Monte Carlo (MC) event generators such as GENIE~\cite{And09}, NuWro~\cite{Gol12}, and NEUT~\cite{Hay21}, where they are combined with other reaction channels to produce complete cross-section predictions. However, rigorous 2p2h calculations require solving the nuclear many-body problem. Ab initio approaches, such as Green’s Function Monte Carlo (GFMC) ~\cite{Lov16}, provide reasonable agreement with electron scattering data for light nuclei, but they cannot be applied to heavier nuclei like $^{40}$Ar and are computationally prohibitive for event-by-event Monte Carlo simulations. Consequently, these generators typically implement the 2p2h channel using phenomenological models and precomputed nuclear response tables for a limited set of reference nuclei~\cite{Dol19}.

In practice, MC generators usually adopt symmetric nuclei (e.g., $^{12}$C or $^{16}$O) as references and extrapolate their 2p2h responses to heavier and asymmetric systems such as $^{40}$Ar or $^{208}$Pb using ad hoc scaling formulas~\cite{Dol19,Sch16,Mos16b}. These formulas often rely on simple dependencies on the Fermi momentum and/or nucleon number. The main limitation is that these dependencies lack a solid theoretical foundation, and different models employ distinct scaling prescriptions. Using a scaling formula to obtain the response of $^{40}$Ar from that of $^{40}$Ca or $^{12}$C introduces an uncertainty; since it has never been benchmarked against exact calculations, its validity-and magnitude of the uncertainty due to neutron-proton asymmetry-remains unknown.

In this work, we address this problem by performing, as far as we know, the first calculation of 2p2h responses in asymmetric nuclear matter. This is achieved by assigning different Fermi momenta to neutrons and protons, following an approach similar to that of Ref.~\cite{Bar18} for 1p-1h processes. We extend our previous study on $^{12}$C by explicitly accounting for the neutron-proton imbalance, and compare $^{40}$Ar with the isoscalar $^{40}$Ca in order to identify the effects of asymmetry. To further assess the role of nuclear asymmetry, we also examine whether the 2p2h response of $^{40}$Ar can be reliably obtained by scaling from the responses of the reference nuclei.

The structure of the paper is as follows. In Sect.~\ref{section2} we develop the relativistic framework for two-particle-two-hole  meson-exchange currents in asymmetric nuclei, introducing separate Fermi momenta for protons and neutrons. Sect.~\ref{section3} presents a detailed comparison of 2p2h response functions for $^{12}$C, $^{40}$Ar, and $^{40}$Ca. In Sect.~\ref{section4} we study the scaling properties of the 2p2h responses and present formulas to predict responses in arbitrary nuclei based on reference calculations. Sect.~\ref{section5} applies these results to describe inclusive electron and neutrino scattering cross sections, comparing with experimental data. Finally, Sect.~\ref{section6} summarizes our conclusions and outlines future developments.
 ------------------------------------------------------------
\section{Relativistic 2p2h framework for asymmetric nuclei}\label{section2}
In this section, we present the formalism for 2p2h processes in asymmetric nuclei. We calculate the two-particle-two-hole meson-exchange current contributions within a relativistic framework, based on the relativistic mean-field (RMF) approach for nuclear matter~\cite{Ros80,Ser86,Weh93}. In the RMF model, nucleons are described as Dirac spinors that interact through mean scalar and vector fields, which effectively modify their propagation inside the nucleus. The energy of a nucleon with momentum $h$ is given by
\begin{equation}
E = \sqrt{h^2 + (m_N^*)^2} + E_v,
\end{equation}
where $m_N^*$ is the nucleon effective mass that accounts for the attractive scalar potential, and $E_v$ is the vector energy due to the repulsive vector field.

The 2p2h excitations arise from two-body current operators acting on nucleon pairs. The model of inclusive scattering in the relativistic fermi gas (RFG) was developed in Ref.~\cite{Rui17}, and later modified to account for the RMF framework~\cite{Mar21a}. Here, we further generalize the expression for the hadronic tensor to make it applicable to systems with different numbers of protons and neutrons. To achieve this, we introduce different Fermi momenta, following the approach proposed in Ref.~\cite{Bar18}. The volume of the initial system is related to the particle numbers and the corresponding Fermi momenta by
\begin{equation}
\frac{V}{(2\pi)^3}
= \frac{3\,Z}{8\pi\,k_{Fp}^{3}}
= \frac{3\,N}{8\pi\,k_{Fn}^{3}},
\label{volume}
\end{equation}
where $Z$ is the proton number, $N$ is the neutron number, and $k_{Fp}$ ($k_{Fn}$) denotes the proton (neutron) Fermi momentum. The hadronic tensor for a two-nucleon emission channel reads
\begin{align}
W^{\mu\nu}_{N'_1 N'_2} = & \frac{V}{(2\pi)^9} 
\int 
d^3p'_1 
d^3p'_2 
d^3h_1 
d^3h_2  
\frac{(m_N^*)^4}{E_1 E_2 E'_1 E'_2} \;
w^{\mu\nu}_{N'_1N'_2}(\mathbf{p}'_1,\mathbf{p}'_2,\mathbf{h}_1,\mathbf{h}_2) \nonumber \\
& \times \Theta_{N'_1,N_1}(p'_1,h_1) \, \Theta_{N'_2,N_2}(p'_2,h_2) \delta(E'_1 + E'_2 - E_1 - E_2 - \omega)  \delta(\mathbf{p}'_1 + \mathbf{p}'_2 - \mathbf{h}_1 - \mathbf{h}_2 - \mathbf{q}),
\label{hadronic12}
\end{align}
where $N'_1$ and $N'_2$ denote the final nucleons, and the indices $N_1, N_2$ denote the initial nucleons isospins. The Pauli-blocking operator for a particle-hole pair of isospin $(N'_i, N_i)$ is given by

\begin{equation}
\Theta_{N'_i,N_i}(p',h)
= \theta (p'-k_{F N'_i} )\,
  \theta (k_{F N_i}-h ),
\end{equation}
where $N_i, N'_i$ can be proton or neutron (p or n), $p'$ is the particle momentum above the Fermi sea with isospin $N'_i$, and $h$ is the hole momentum below the Fermi sea with isospin $N_i$. The function $\theta(x)$ denotes the Heaviside step function.

The elementary tensor, $w^{\mu\nu}_{N'_1N'_2}$, for a nucleon pair is constructed from antisymmetrized two-body current matrix elements. 
\begin{equation}
w^{\mu\nu}_{N'_1N'_2} = G_s \sum_{s_1,s_2,s'_1,s'_2}
  j^\mu_A(1',2',1,2)^*\, j^\nu_A(1',2',1,2).
\end{equation}
Here $s_i$ and $s'_i$ label the initial and final spin projections, respectively, and the transition is denoted as $\lvert 1,2 \rangle \to \lvert 1',2' \rangle$. The sums run over the spin states of the initial and final nucleons, and $j^\mu_A$ denotes the antisymmetrized two-body current with respect to identical particles. The factor $G_s$ enforces the antisymmetry of the two-nucleon state under exchange of the two particles---equivalently, under interchange of their momenta and spin quantum numbers---and prevents double counting of identical-particle contributions. In inclusive scattering, $G_s = 1/2$ for neutrino scattering, $G_s = 1/4$ for the $pp$ and $nn$ channels in electron scattering, and $G_s = 1$ for the $np$ channel in electron scattering in the final state.

The two-body current operators $j^\mu$ used in this work follow the expressions derived the relativistic expressions derived in Refs~\cite{Her07,Rui17}. They are subsequently modified, as detailed in Eqs.~(28)-(32) of Ref.~\cite{Mar21a}, to incorporate the nucleon effective mass within the RMF framework. In addition, we include in-medium modifications of the $\Delta$ resonance, assuming a universal coupling as in Ref.~\cite{Mar23b}.

It is important to note that, rather than directly evaluating the full inclusive hadronic tensor, Eq.~(\ref{hadronic12}), we follow the standard approach of working with nuclear response functions. In this framework, the integration over $\np'_2$ is carried out using the three-momentum $\delta$-function. The nuclear response function, expressed in terms of two Fermi momenta, can be written in terms of the reduced response function for a nucleon pair, $r^K$, as
\begin{eqnarray}
R^{K}_{N'_2 N'_2}
= &&
\frac{V}{(2\pi)^9}
\int d^3p'_1 
 d^3h_1 
  d^3h_2  \frac{(m_N^*)^4}{E_1E_2E'_1E'_2} r^{K}(\np'_1,\np'_2,\nh_1,\nh_2) \Theta_{N'_1,N_1}(p'_1,h_1)\Theta_{N'_2,N_2}(p'_2,h_2)\nonumber \\
&& \times \delta(E'_1+E'_2-E_1-E_2-\omega) ,
\label{responses}
\end{eqnarray}

We exploit the axial symmetry around the $z$-axis, chosen along the momentum transfer $\mathbf{q}$. In particular, the azimuthal angle of the first outgoing particle can be fixed to $\phi'_1 = 0$, with an overall factor of $2\pi$ included to account for this symmetry. Furthermore, the energy $\delta$-function enables us to analytically perform the integration over $p'_1$, thereby reducing Eq.~(\ref{hadronic12}) to a seven-dimensional integral. This reduced integral is then evaluated numerically, either in the laboratory frame~\cite{Sim14} or in the two-hole center-of-mass frame~\cite{Mar23a, Mar23b}.
 
For electron scattering ($K = L_{\text{em}}, T_{\text{em}}$), only the longitudinal and transverse components ($R^L_{\text{em}}$, $R^T_{\text{em}}$) contribute.
\begin{align}
R^L_{\text{em}} &= W^{00}_{\text{em}}  \\
R^T_{\text{em}} &= W^{11}_{\text{em}} + W^{22}_{\text{em}}
\end{align}

For neutrino and antineutrino scattering ($K = \text{CC}, \text{CL},,\text{LL}, T, T'$), five response functions become relevant: $R^{CC}$, $R^{CL}$, $R^{LL}$, $R^T$, and $R^{T'}$.
The standard components are the next lineal combination:
\begin{align}
R^{CC}   &= W^{00}  \\
R^{CL}   &= -\frac{1}{2} \left( W^{03} + W^{30} \right)  \\
R^{LL}   &= W^{33} \\
R^{T}    &= W^{11} + W^{22}  \\
R^{T'}   &= -\frac{i}{2} \left( W^{12} - W^{21} \right) 
\end{align}

\section{Comparison of 2p2h responses between $^{12}$C, $^{40}$Ar, and $^{40}$Ca}\label{section3}
In this section, we compare the 2p2h nuclear response functions for $^{12}$C, $^{40}$Ar, and $^{40}$Ca within two models, RMF and RFG. The model setups are as follows. In the RMF calculation we employ in-medium
effective masses $m_N^*$ and $m_\Delta^*$ (i.e., $M_N^* \equiv m_N^*/m_N < 1$, $M_\Delta^* \equiv m_\Delta^*/m_\Delta < 1$) and we do not introduce a phenomenological separation energy. In the RFG calculation we instead use free masses
($M_N^*=M_\Delta^*=1$) and include a separation energy $E_s$, entering the lepton kinematics through $\omega'=\omega+E_s$. In both frameworks we allow for distinct proton and neutron Fermi momenta $(k_{Fp},k_{Fn})$ to account for isospin asymmetry. The parameters used in this work are summarized in Table~\ref{tablanuclei}.

The effective mass of the $\Delta$, $m_\Delta^*$, enters only through the two-body MEC currents, specifically in the $\Delta$-forward and $\Delta$-backward contributions, via the $\Delta$ propagator that appears in those terms. It does not affect the single-nucleon kinematics, the phase-space factors, or Pauli blocking. For the $\Delta$ current, the RMF calculation employs the full in-medium $\Delta$ propagator (including both real and imaginary parts), whereas in the RFG framework only the real part of the free propagator is retained. The RFG is the same framework used in the SuSAv2+MEC model~\cite{Meg16}, which has been implemented in Monte Carlo event generators.

The numerical evaluation of the full seven-dimensional integral in Eq.~(\ref{responses})
was carried out on the Proteus high-performance computing cluster~\cite{proteus}.
We resolve contributions by the isospin of the final state
nucleon pair: for neutrino-induced reactions we consider $pp$ and $np$ final pairs; for
electron scattering we additionally include $nn$ pairs.

\begin{figure}[htp]
\centering
\includegraphics[scale=0.85, bb=40 405 550 780]{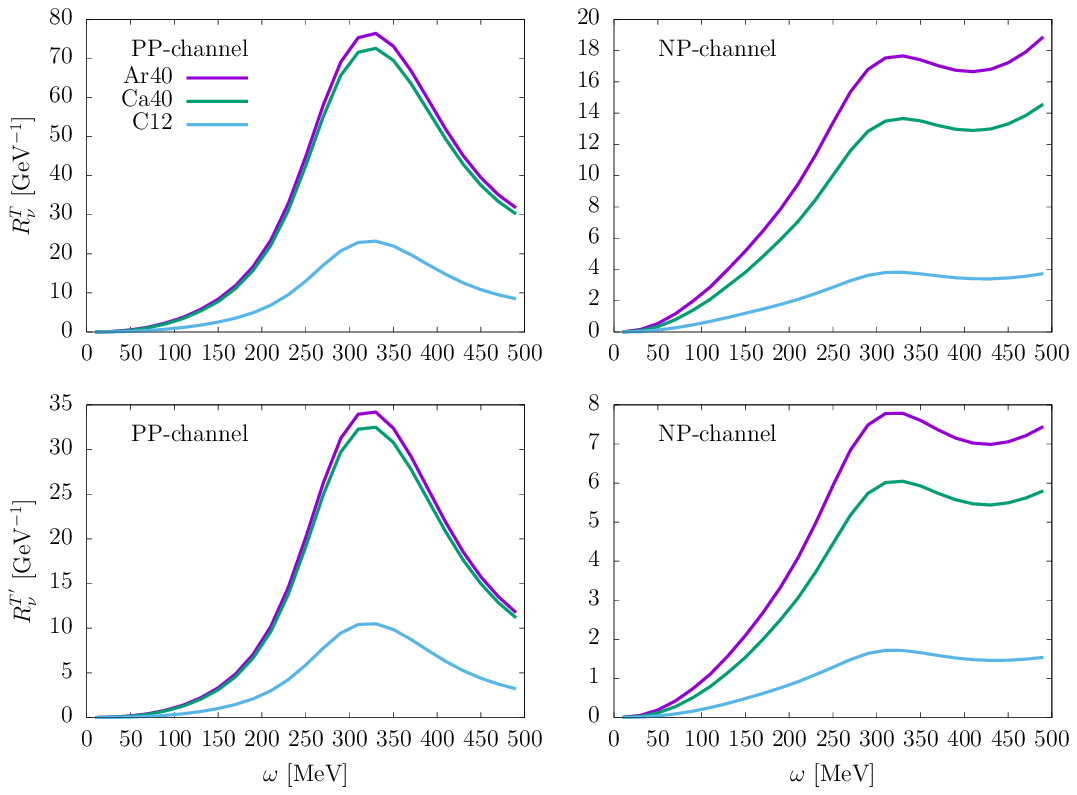}
\caption{RMF calculation of 2p2h responses for charged-current neutrino scattering at $q=500~\text{MeV}/c$ in the $pp$ and $np$ channels, shown for $^{40}$Ar, $^{40}$Ca, and $^{12}$C.}
\label{fig1rmf}
\end{figure}

Figure~\ref{fig1rmf} shows the neutrino $R^T$ and $R^{T'}$ responses obtained within the RMF calculation for $q = 500$ MeV/c in the three nuclei considered. In the $pp$ channel (left panels), the maximum in the $\Delta$-resonance region appears at $\omega \simeq 300$-$350$ MeV, with a pronounced peak. In the $np$ channel (right panels), a peak is also present at the same energy transfer, but it is noticeably less pronounced. This behavior arises because, in the $np$ channel, the $\Delta$-backward diagram dominates because a cancellation occurs in the $\Delta$-forward contribution~\cite{Mar24b}. For these kinematics, the $pp$ channel strength is more than four times larger than that of the $np$ channel, whose response is smaller and varies more smoothly with $\omega$. A direct comparison between $^{40}$Ar and $^{40}$Ca---two nuclei with same mass numbers but different numbers of protons and neutrons---shows that the $pp$ responses are very similar across the entire $\omega$ range. For $^{40}\mathrm{Ar}$, the pp responses are modestly larger than those of $^{40}\mathrm{Ca}$, by about 6\%, whereas the np responses exceed those of $^{40}$Ca by more than 30\%.

\begin{table}[htb]
\centering
\begin{ruledtabular}
\begin{tabular}{cccccccc}
Nucleus & $Z$ & $N$ & $k_{Fp}$ & $k_{Fn}$ & $M_N^*$ & $M_\Delta^*$ & $E_s$ \\
\hline
$^{12}$C  & 6  & 6  & 225 & 225 & 0.80 & 0.845 & 40 \\
$^{40}$Ar & 18 & 22 & 237 & 256 & 0.73 & 0.79  & 56 \\
$^{40}$Ca & 20 & 20 & 240 & 240 & 0.73 & 0.79  & 56
\end{tabular}
\end{ruledtabular}
\caption{Parameters used in the calculations. Fermi momenta $k_{Fp}$ and $k_{Fn}$ are in
MeV/$c$; $E_s$ is the separation energy in MeV. The effective-mass ratios $M_N^*$ and
$M_\Delta^*$ are used in the RMF calculation. In the RFG calculation we set
$M_N^*=M_\Delta^*=1$ and use the listed $E_s$.}
\label{tablanuclei}
\end{table}

\begin{figure}[htp]
\centering
\includegraphics[scale=0.85, bb=40 405 550 780]{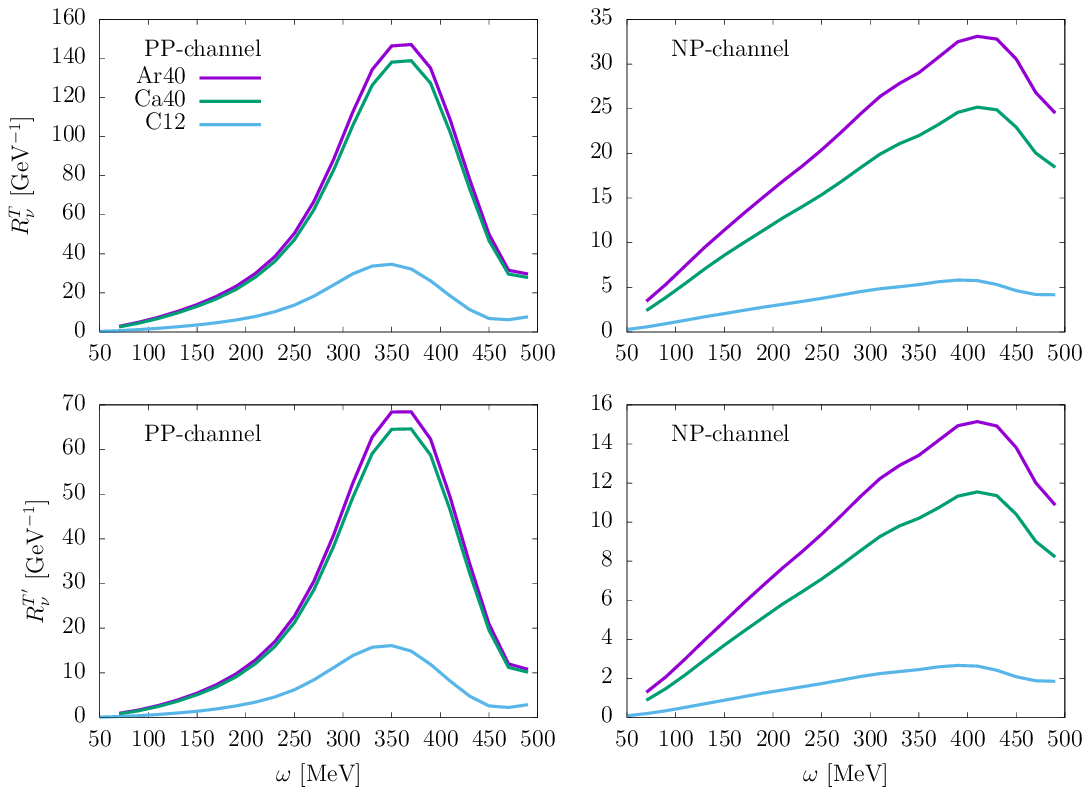}
\caption{The same as Fig.~\ref{fig1rmf}, but computed in the RFG
framework (free masses, with separation energy, and real part of the free $\Delta$
propagator only).}
\label{fig2rfg}
\end{figure}

The corresponding RFG calculation is shown in Fig.~\ref{fig2rfg}. Relative to the RMF,
the RFG responses are typically larger and the $\Delta$ peak is narrower and shifted by approximately $40$~MeV. These
differences arise from (i) the treatment of the $\Delta$ current (full in-medium propagator
in RMF versus real part of the free propagator in RFG), which impacts the peak position and
width, and (ii) the use of effective masses ($m_N^*$ and $m_\Delta^*$) in RMF as opposed to free masses plus
a separation energy in RFG.
The same trend persists in the RFG model: the pp responses of $^{40}\mathrm{Ar}$ remain slightly larger than those of $^{40}\mathrm{Ca}$, while the difference is much more pronounced for np emission.

\begin{figure}[htp]
\centering
\includegraphics[scale=0.85, bb=16 390 563 775]{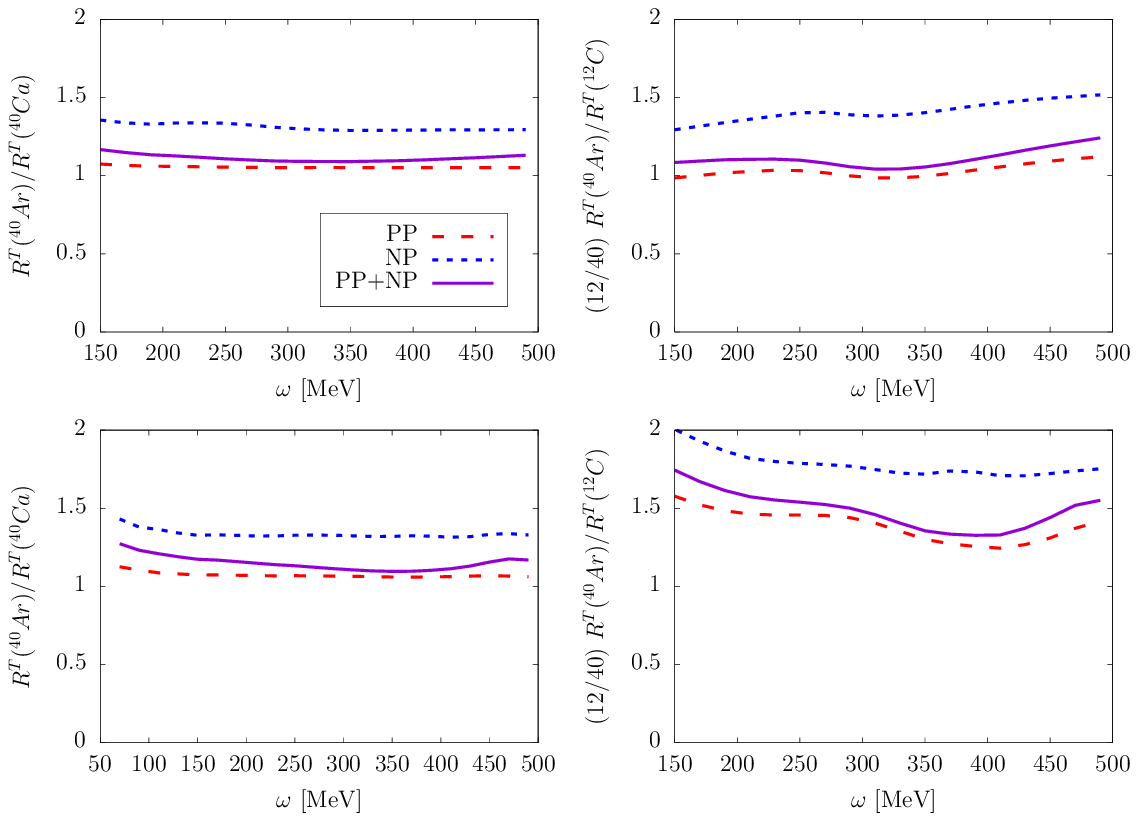}
\caption{Ratio of the $^{40}$Ar transverse neutrino response to that of $^{40}$Ca (left panels) and $^{12}$C (right panels) in the $pp$, $np$, and $pp+np$ emission channels at $q = 500$~MeV/$c$. Upper panels: RMF model; lower panels: RFG model.}
\label{figura3}
\end{figure}

Regardless of whether the RMF or RFG framework is used, the 2p2h-MEC $R^T$ and $R^{T'}$ responses of $^{12}\mathrm{C}$, $^{40}\mathrm{Ca}$, and $^{40}\mathrm{Ar}$ exhibit very similar shapes, differing primarily by an overall scaling factor. To quantify this, Fig.~\ref{figura3} shows the per-nucleon ratio of $^{40}\mathrm{Ar}$ to $^{40}\mathrm{Ca}$ (left panels) and to $^{12}\mathrm{C}$ (right panels), for both RMF (top row) and RFG (bottom row) calculations. At $q = 500~\mathrm{MeV}/c$, these ratios vary smoothly with the energy transfer-lying within the range $[1,2]$-for both the $pp$ and $np$ channels in both models. This indicates a weak dependence on the energy transfer and supports the use of an approximately constant scaling factor.

The next step is to examine whether the per-nucleon ratio between the responses of different nuclei remains stable under variations of the momentum transfer $q$. This expectation is supported both by the analytic arguments of Ref.~\cite{Ama17}, which suggest that 2p2h responses in different nuclei scale by an approximately constant factor, and by the scaling prescription of the Valencia model implemented in GENIE~\cite{Sch16}, which rescales according to the number of available nucleon pairs.

\begin{figure}[htp]
\centering
\includegraphics[scale=0.85, bb=160 390 413 775]{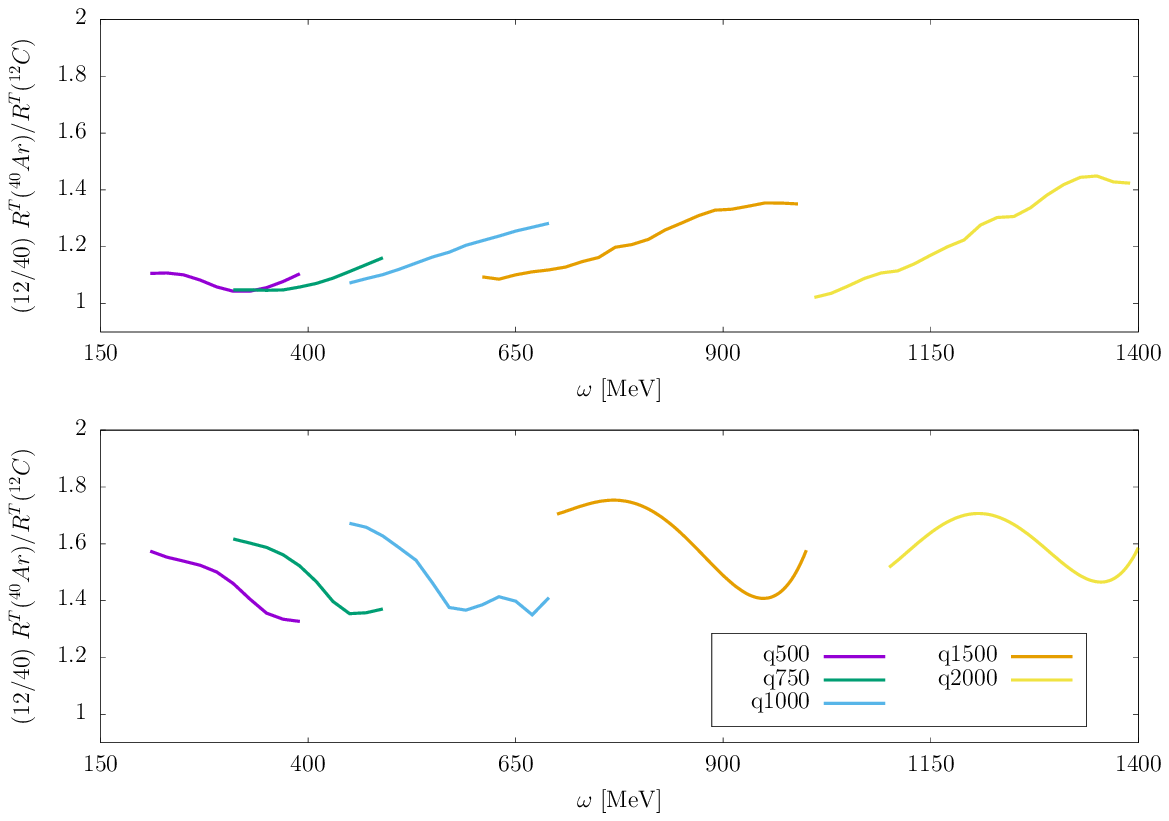}
\caption{Ratio of the 2p2h MEC transverse neutrino response of $^{40}$Ar to that of $^{12}$C in the $pp+np$ emission channel for several momentum transfers ($q=500$, 750, 1000, 1500, 2000~MeV/$c$). Upper panels: RMF model; lower panels: RFG model.}
\label{figura4}
\end{figure}

Figure~\ref{figura4} shows the per-nucleon ratio $^{40}\mathrm{Ar}/^{12}\mathrm{C}$ for several momentum transfers in both RMF and RFG. The curves are smooth, exhibiting only a mild dependence on $\omega$. The density-scaling estimate of Ref.~\cite{Ama17}, proportional to $Z ~ k_F^2$, predicts a per-nucleon ratio between 1.00 and 1.16 (depending on the value of the Fermi momentum used) for $^{40}\mathrm{Ar}$ in both models. In contrast, our calculations yield characteristic per-nucleon ratios of $\approx 1.1$ in RMF and $\approx 1.5$ in RFG. This discrepancy arises because the density scaling formula neglects proton-neutron asymmetry ($k_{Fp}\neq k_{Fn}$) and separation-energy effects, thereby underestimating the Argon strength, particularly in the RFG case.

Table \ref{tabla3} reports the per-nucleon ratios \(R^T_{\nu}(^{40}\mathrm{Ar})/R^T_{\nu}(^{12}\mathrm{C})\) evaluated at the maximum of the \(\Delta\)-MEC resonance in the total channel \(pp{+}pn\) in neutrino scattering for RMF and RFG framework, covering \(q=500,\,750,\,1000,\,1500\), and \(2000\)~MeV/\(c\). Although we find a kinematic dependence (q, $\omega$) in the ratio of the 2p2h responses for different nuclei, this dependence is relatively mild in the region that contributes most to the cross section, and therefore it is of practical interest to implement an approximate scaling formula that allows one to estimate the response of $^{40}$
Ar from that of $^{12}$C.

\begin{table}[h!]
\centering
\begin{ruledtabular}
\begin{tabular}{ccc}
$q$ (MeV/$c$) & RMF ($pp{+}pn$) & RFG ($pp{+}pn$) \\
\hline
500  & 1.04 & 1.33 \\
750  & 1.10 & 1.37 \\
1000 & 1.16 & 1.39 \\
1500 & 1.28 & 1.49 \\
2000 & 1.30 & 1.53
\end{tabular}
\end{ruledtabular}
\caption{Per-nucleon ratio $R^T_{\nu}(^{40}\mathrm{Ar})/R^T_{\nu}(^{12}\mathrm{C})$ at the $\Delta$-peak maximum for each $q$ in the $pp{+}pn$ channel, for RMF and RFG.}
\label{tabla3}
\end{table}

\section{2p2h scaling properties}\label{section4}
In this section, we propose asymmetric scaling formulas for 2p2h nuclear responses that enable prediction for an arbitrary nucleus starting from a single reference calculation-in our case, $^{12}$C. The purpose is not merely to refine the description currently implemented in Monte Carlo event generators for asymmetric nuclei, but to provide a general and useful scaling prescription for each emission channel, so that responses for different targets can be obtained without repeating computationally expensive microscopic calculations. Our formulation rests on two empirical observations emerging from fully microscopic RMF and RFG calculations. 
First, the ratio of 2p2h responses between two nuclei is nearly constant over a broad kinematic range. For any two nuclei $X$ and $Y$, we assume:
\begin{equation}
\frac{R^{K}(X;q,\omega)}{R^{K}(Y;q,\omega)} \approx \text{const.}\,,
\label{ratio}
\end{equation}
where $K = CC, CL, LL, T, T', L_{\rm em}, T_{\rm em}$ denotes a given response component. Second, this near-constant ratio is primarily governed by phase-space effects, determined by the proton and neutron numbers and the Fermi momenta of each nucleus. The responses can be expressed in terms of the averaged single-pair responses, $\langle r^{K}_{N_1N_2}\rangle$, as:
\begin{equation}
R^{K}_{\mathrm{N'_1 N'_2}}(X;q,\omega) \propto V F_{N_1N_2}(q,\omega;X) \frac{\langle r^{K}_{N'_1N'_2}(q,\omega)\rangle}{(2\pi)^9}\,,
\end{equation}
where $F_{N_1N_2}$ involves the kinematic dependence of the 2p2h phase space.
Within the frozen-nucleon approximation~\cite{Sim17b}, which neglects initial nucleon momenta compared to $q$ and it is accurate for except very near threshold, the phase-space factor can be evaluated analytically as
\begin{equation}
F_{N_1N_2}(q,\omega;X) = [k_{F N_1}(X)]^{3} [k_{F N_2}(X)]^{3} [m_N^*(X)]^{2}
\sqrt{1 - \frac{4\,m_N^{*2}(X)}{(2 m_N^*(X) + \omega)^{2} - q^{2}}}\,.
\label{eq:frozen_scaling}
\end{equation}
This implies a leading dependence proportional to $[k_{F N_1}]^{3}[k_{F N_2}]^{3}$. Moreover, the volume factor satisfies $V/(2\pi)^3 \propto Z/k_{F p}^{3} \propto N/k_{F n}^{3}$ [cf. Eq.~(\ref{volume})], so that, schematically, the leading behavior can be summarized as the following 2p2h asymmetric scaling rule:

\begin{equation} 
R^{K}_{N'_1N'_2}(X)\;\propto\;Z (X)\; [k_{Fp}(X)]^{-3} [k_{FN_1}(X)]^3 [k_{FN_2}(X)]^3 \, 
\label{relacion} 
\end{equation}

This formula exhibits similar dependencies to the semi-empirical formulas previously developed in Refs.~\cite{Mar21b,Mar23a}. Since our interest lies in ratios rather than absolute responses, the dependence on the Fermi momentum and the proton number factorizes, while the remaining nuclear effects are contained in a quantity referred to as the reduced coefficients. We therefore write explicit asymmetric scaling formulas for a generic nucleus $X$, using ${}^{12}\mathrm{C}$ as a reference, distinguishing between electromagnetic and neutrino responses and among the $pp$, $np$, and $nn$ final-state channels. Here, $Z(X)$ and $N(X)$ denote the proton and neutron numbers of nucleus $X$.

Scaling formulas for 2p2h responses in electron scattering
\begin{align}
R^K_{pp}(X) &=
R^K_{pp}(^{12}\mathrm{C})\;
\frac{Z(X)}{Z(^{12}\mathrm{C})}\,
\left[\frac{k_{Fp}(X)}{k_{Fp}(^{12}\mathrm{C})}\right]^{3}
\mathcal{C}^\mathrm{ em}_{pp}(X),\label{sfe}
\\[4pt]
R^K_{np}(X) &=
R^K_{np}(^{12}\mathrm{C})\;
\frac{Z(X)}{Z(^{12}\mathrm{C})}\,
\left[\frac{k_{Fn}(X)}{k_{Fn}(^{12}\mathrm{C})}\right]^{3}
\mathcal{C}^\mathrm{ em}_{np}(X),
\\[4pt]
R^K_{nn}(X) &=
R^K_{nn}(^{12}\mathrm{C})\;
\frac{N(X)}{N(^{12}\mathrm{C})}\,
\left[\frac{k_{Fn}(X)}{k_{Fn}(^{12}\mathrm{C})}\right]^{3}
\mathcal{C}^\mathrm{em}_{nn}(X).
\end{align}

Scaling formulas for 2p2h responses in neutrino scattering
\begin{align}
R^K_{pp}(X) &=
R^K_{pp}(^{12}\mathrm{C})\;
\frac{Z(X)}{Z(^{12}\mathrm{C})}\,
\left[\frac{k_{Fn}(X)}{k_{Fn}(^{12}\mathrm{C})}\right]^{3}
\mathcal{C}^\nu_{pp}(X),
\\[4pt]
R^K_{np}(X) &=
R^K_{np}(^{12}\mathrm{C})\;
\frac{N(X)}{N(^{12}\mathrm{C})}\,
\left[\frac{k_{Fn}(X)}{k_{Fn}(^{12}\mathrm{C})}\right]^{3}
\mathcal{C}^\nu_{np}(X).
\label{sfnu}
\end{align}

In general, the behavior of these scaling formulas for the 2p2h responses reflects the phase space of the initial nucleons participating in the interaction. Each channel ($pp$, $np$, $nn$) is therefore weighted both by the number of protons or neutrons available in the initial state and by the cubic power of the corresponding Fermi momentum. After cancellation with the volume factor, the result retains the dependence on $Z$, $N$, and on $k_{Fp}^3$ and $k_{Fn}^3$. For example, in neutrino scattering, the final $pp$ channel scales with $k_{Fn}$, because the phase space depends on the Fermi momenta of both protons and neutrons, but the $k_{Fp}^3$ factor cancels out, leaving only the dependence on $k_{Fn}^3$ and the number of protons Z. Similar results are obtained for the other channels.

The computed reduced coefficients $\mathcal{C}_{N'_1N'_2}(X)$ are listed in Table~\ref{tablaprop} for both RFG and RMF and for the different emission channels. By construction, these coefficients are dimensionless, nucleus- and model-dependent constants that are independent of $q$ and $\omega$, and common to all response components $K$. Recall that these factors encapsulate residual nuclear effects beyond the phase-space scaling. In the RMF case, this includes the dependence on effective parameters of the model such as the Fermi momentum and the effective masses, which have been fixed in previous works (see, e.g., ~\cite{Mar17,Ama18,Mar21a}). Similarly, in the RFG approach, the separation energy is incorporated through the corresponding effective parameter.

\begin{table}[h!]
\centering
\begin{ruledtabular}
\begin{tabular}{ccccccc}
\multirow{2}{*}{Model} & \multirow{2}{*}{Nucleus} & \multicolumn{2}{c}{Neutrino ($\nu$)} & \multicolumn{3}{c}{Electron (em)} \\
 & & $\nu\,pp$ & $\nu\,np$ & $e\,pp$ & $e\,np$ & $e\,nn$ \\
\hline
\multirow{3}{*}{RFG}
 & $^{12}$C & 1.00 & 1.00 & 1.00 & 1.00 & 1.00 \\
 & $^{40}$Ar & 0.98 & 1.08 & 1.10 & 1.06 & 1.11 \\
 & $^{40}$Ca & 0.99 & 1.08 & 1.10 & 1.06 & 1.09 \\
\hline
\multirow{3}{*}{RMF}
 & $^{12}$C & 1.00 & 1.00 & 1.00 & 1.00 & 1.00 \\
 & $^{40}$Ar & 0.76 & 0.87 & 0.85 & 0.76 & 0.87 \\
 & $^{40}$Ca & 0.78 & 0.88 & 0.87 & 0.78 & 0.87
\end{tabular}
\end{ruledtabular}
\caption{Computed scaling coefficients $\mathcal{C}_{N'_1N'_2}$ for $^{40}$Ar and $^{40}$Ca in different channels, normalized to $^{12}$C.}
\label{tablaprop}
\end{table}

\begin{figure}[htp]
\centering
\includegraphics[scale=0.85, bb=41 195 560 780]{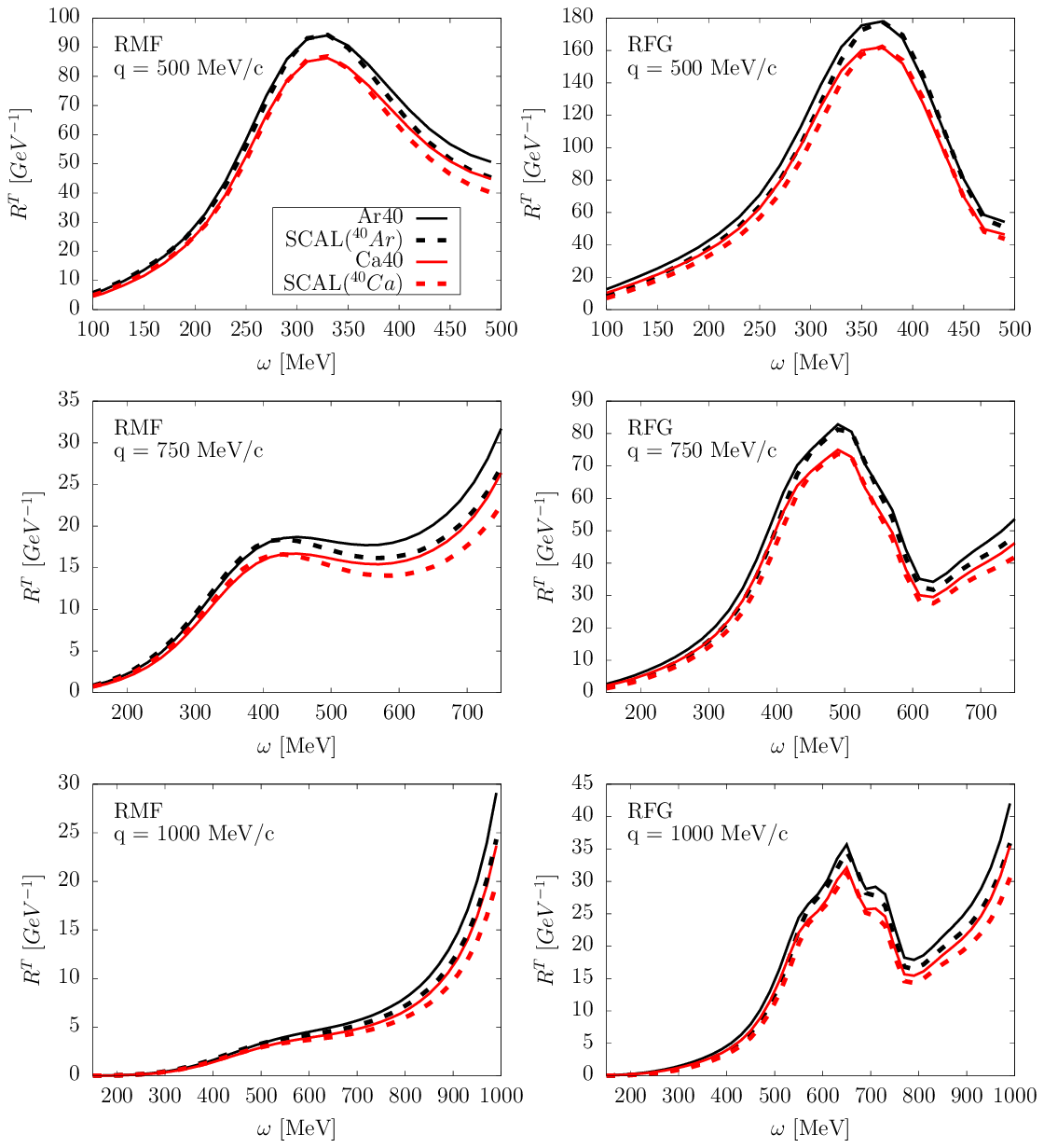}
\caption{Transverse neutrino response in the $pp+pn$ channel. The exact responses for $^{40}$Ar (black) and $^{40}$Ca (red) are compared with the predictions obtained by rescaling from $^{12}$C using the scaling formulas with the parameters of Table~\ref{tablaprop}, shown as dashed lines of the same color as the target nucleus. Results are displayed for $q=500$, $750$, and $1000$~MeV/$c$ in both the RMF (left column) and RFG (right column) frameworks.}

\label{figura5}
\end{figure}

The performance of the scaling procedure, based on the asymmetric scaling formulas Eqs.~(\ref{sfe}-\ref{sfnu}) and the parameters in Table~\ref{tablaprop}, is illustrated in Fig.~\ref{figura5}, where we compare the neutrino transverse response in the total $pp{+}np$ channel at three momentum transfers ($q = 500$, $750$, and $1000\mathrm{MeV}/c$). The left column shows RMF results, while the right column shows RFG results. For each value of $q$, the exact ${}^{40}\mathrm{Ar}$ response (black) and the exact ${}^{40}\mathrm{Ca}$ response (red) are shown, along with the prediction obtained by scaling from ${}^{12}\mathrm{C}$, indicated by dashed lines of the same color as the target nucleus, using the scaling prescription and the neutrino reduced coefficients $\mathcal{C}^\nu_{N_1N_2}$ from Table~\ref{tablaprop}. Two main trends emerge:

\begin{itemize}
\item The exact ${}^{40}\mathrm{Ca}$ result provides a close proxy to ${}^{40}\mathrm{Ar}$ in the $pp + np$ channel, although the difference in the $np$ channel is larger than in the $pp$ channel, as observed in Figs.~\ref{fig1rmf} and \ref{fig2rfg}; this is because the $np$ response is smaller than the $pp$ response.
\item The present scaling reproduces the exact ${}^{40}\mathrm{Ar}$ and ${}^{40}\mathrm{Ca}$ responses across a wide range of $\omega$.
\end{itemize}

\begin{figure}[htp]
\centering
\includegraphics[scale=0.85, bb=41 595 560 780]{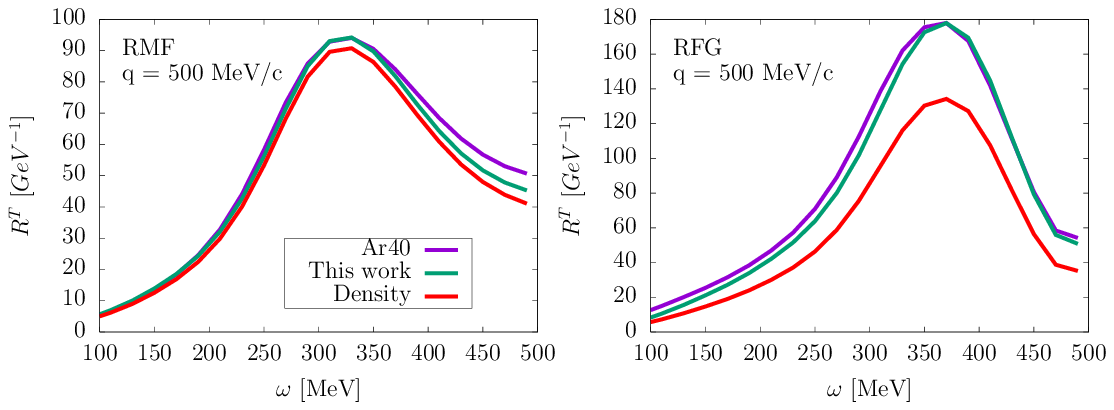}
\caption{Transverse 2p2h neutrino response in $^{40}$Ar at $q=500$~MeV/$c$ in RMF and RFG. The response rescaled from $^{12}$C using the presented formula is compared with the density-based prescription of Ref.~\cite{Ama17} (labeled Density) and the exact microscopic calculation.}
\label{figura6}
\end{figure}

Finally, Fig.~\ref{figura6} presents a comparison between the scaling formulas proposed in this work and the density-based prescription of Ref.~\cite{Ama17}, which is implemented in GENIE. For this comparison, we start from the exact ${}^{12}\mathrm{C}$ results and rescale them to obtain ${}^{40}\mathrm{Ar}$. In the density-based approach, this corresponds to multiplying by $Z k_F^2$ of ${}^{40}\mathrm{Ar}$ and dividing by the corresponding values for ${}^{12}\mathrm{C}$. Here, we adopt an average Fermi momentum of $k_F = 240~\mathrm{MeV}/c$ for Argon, intermediate between the proton and neutron values. In the figure, the density-based prescription is labeled as Density. In contrast, our asymmetric scaling prescription, given by Eqs.~(\ref{sfe}-\ref{sfnu}), treats each reaction channel separately, applying the corresponding reduced coefficients listed in Table~\ref{tablaprop}.

The results show that the density-based formula performs reasonably well within the RMF framework; however, in the RFG case for Argon the deviations are significantly larger, reaching up to $\sim$38\%. Moreover, our exact calculations indicate that the ${}^{40}\mathrm{Ar}$ cross section exceeds that of ${}^{40}\mathrm{Ca}$, whereas the density prescription-being proportional to $Z$ and not distinguishing between protons and neutrons- predicts a smaller value. This underscores the need for an improved asymmetric scaling scheme, such as the one presented here.

\section{Cross Sections}\label{section5}
In this section we calculate the electron and neutrino cross section for different nuclei and compare with available experimental data, including the various contributions that are relevant to the total cross section. To this end, we employ the asymmetric scaling formulas introduced in the previous sections to evaluate, from pre-computed ${}^{12}\mathrm{C}$ response functions, the 2p2h-MEC contribution to the cross section for other nuclei. We present predictions for a recent electron-scattering experiment on a ${}^{12}\mathrm{C}$ target, as well as for ${}^{40}\mathrm{Ar}$. In the case of neutrinos, we compare our theoretical results for ${}^{40}\mathrm{Ca}$ and ${}^{40}\mathrm{Ar}$-both obtained using the scaling formulas-and benchmark the model with the experimental data from the MicroBooNE measurement on a ${}^{40}\mathrm{Ar}$ target.

\subsection{Inclusive Electron Scattering}
We begin by considering inclusive electron scattering off nuclei, where the double-differential cross section is given by
\begin{equation}
\frac{d\sigma}{d\Omega d\epsilon'}
= \sigma_{\rm Mott}
\left(
v_L R_{em}^L +  v_T  R_{em}^T
\right),
\end{equation}
where $\epsilon$ and $\Omega$ denote the final electron energy and the scattered solid angle, $\omega$ is the energy transfer, $q=|\mathbf{q}|$ the three-momentum transfer, and $Q^2=\omega^2-q^2<0$. The Mott cross section is  $\sigma_{Mott}$ . \\
The leptonic kinematic factors are taken in the standard form:
\begin{equation}
v_L=\left(\frac{Q^2}{q^2}\right)^{\!2},
\qquad
v_T=\frac{1}{2}\left|\frac{Q^2}{q^2}\right|+\tan^2\!\frac{\theta_e}{2},
\end{equation}
Our calculations incorporate the dominant reaction mechanisms operating in this kinematic regime. The quasielastic (QE) contribution is evaluated within the recently developed SuSAMv2 (Superscaling Analysis with Relativistic Effective Mass version 2) framework~\cite{Mar25a}, which represents a substantial advance over the earlier version~\cite{Mar17,Ama18}. In SuSAMv2, the analysis goes beyond a global fit to the inclusive cross section: the longitudinal nuclear response is explicitly constrained by experimental data, while the transverse response is obtained from a separate fit to electron-scattering cross-section measurements, thereby disentangling and extracting the longitudinal contribution. This methodology leads to different longitudinal and transverse scaling functions that depends also on q, thereby providing a significant enhancement in the description of nuclear responses, particularly in the low-momentum-transfer region.

In addition to the QE contribution, our model incorporates two-particle-two-hole (2p2h) meson-exchange currents (MEC)-the central subject of the present work and discussed in detail in the preceding sections-together with short-range correlations (SRC), evaluated through a semi-empirical parametrization fitted to inclusive electron-scattering data~\cite{Mar23a}, and resonance plus deep inelastic contributions, described within the dynamical coupled-channels model (denoted in this work as DIS)~\cite{jesus}. Although a fully microscopic treatment of final-state interactions (FSI) is not yet implemented, their effects are effectively accounted for through the phenomenological extraction of the scaling functions in the SuSAMv2 framework~\cite{Mar25a}.

\begin{figure}[h]
\centering
\includegraphics[scale=0.85, bb=145 361 440 772]{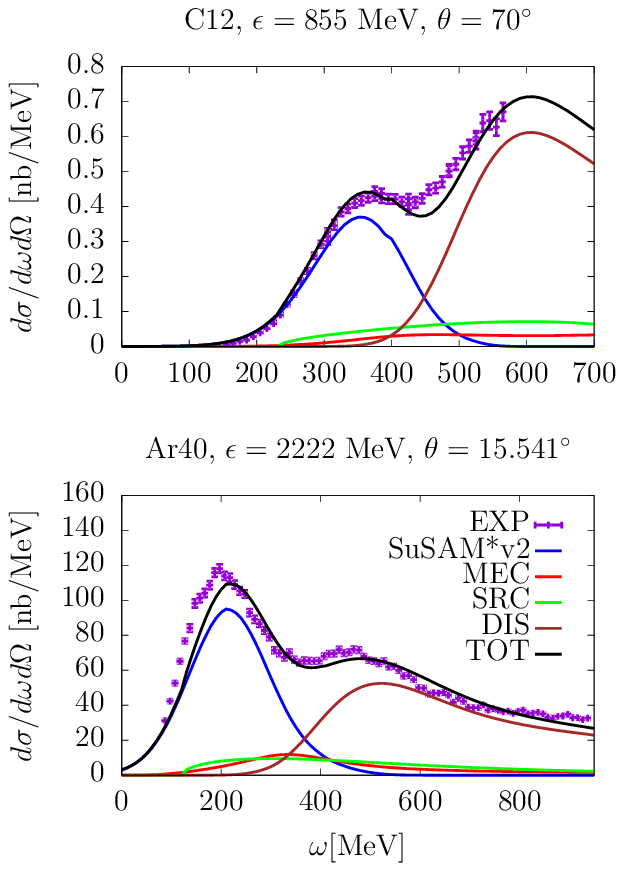}
\caption{Inclusive electron scattering cross sections for $^{12}$C and $^{40}$Ar, including QE, MEC, SRC, and DIS contributions. The $^{40}$Ar response is obtained by applying the rescaling formula to the $^{12}$C calculations. Experimental data are taken from~\cite{archive,archive2,Ben08,Mih24,Dai18}.}
\label{figura7}
\end{figure}

We present illustrative examples of inclusive electron-nucleus cross sections for two representative cases: ${}^{12}$C at high momentum transfer and ${}^{40}$Ar at intermediate momentum transfer. To illustrate the relative importance of the different reaction mechanisms discussed above-quasielastic scattering, two particle-two hole meson exchange currents, short-range correlations , and resonance plus deep---inelastic processes---we show in Figure~\ref{figura7} the individual contributions to the cross section, as well as their sum. The improved SuSAMv2 scaling model provides an accurate description of the quasielastic peak, successfully reproducing its width, position, and height. Both the 2p2h-SRC and 2p2h-MEC mechanisms exhibit similar shapes and primarily contribute in the dip region between the quasielastic and $\Delta$-resonance peaks. At larger energy transfer, beyond the maximum of the 2p2h contributions, pion production and deep-inelastic processes dominate, populating the region above the $\Delta$ resonance. The sum of all these mechanisms yields a reasonable description of the experimental data. The figure for ${}^{40}$Ar is similar with the results obtained by Butkevich et al.~\cite{But20}.

\subsection{Neutrino-Nucleus Scattering}
We now turn our attention to neutrino-induced charged-current quasielastic (CCQE) scattering in nuclei, focusing specifically on the $(\nu_\mu,\mu^-)$ channel. In this process, the incoming neutrino has energy $\epsilon = E_\nu$, and the outgoing muon carries energy $\epsilon' = m_\mu + T_\mu$ and momentum $\vec{k}'$.

The cross section for charged-current quasielastic neutrino scattering, double-differential in muon kinetic energy and scattering angle, takes the following general form:
\begin{eqnarray}
\frac{d^2\sigma}{dT_\mu d\cos\theta_\mu}
&=&
\sigma_0
\left\{
V_{CC} R_{CC}+
2{V}_{CL} R_{CL}
\right.
 \nonumber\\
&&
\left.
+{V}_{LL} R_{LL}+
{V}_{T} R_{T}
\pm
2{V}_{T'} R_{T'}
\right\} \, , 
\end{eqnarray}
where
\begin{equation}
\sigma_0=
\frac{G^2\cos^2\theta_c}{4\pi}
\frac{k'}{\epsilon}v_0,
\end{equation}
with $G = 1.166\times 10^{-11}\ \rm MeV^{-2}$ the Fermi constant, $\theta_c$ the Cabibbo angle ($\cos\theta_c=0.975$), and $v_0 = (\epsilon+\epsilon')^2-q^2$.

The five factors $V_K$ ($K=$ CC, CL, LL, $T$, $T'$) are purely kinematic and do not depend on the details of the nuclear structure. Their explicit forms are:
\begin{align}
V_{CC}  &= 1-\delta^2\frac{Q^2}{v_0}, \\
V_{CL}  &= \frac{\omega}{q}+\frac{\delta^2}{\rho'}\frac{Q^2}{v_0}, \\
V_{LL}  &= \frac{\omega^2}{q^2}+
\left(1+\frac{2\omega}{q\rho'}+\rho\delta^2\right)\delta^2\frac{Q^2}{v_0}, \\
V_{T}   &= \frac{Q^2}{v_0}
+\frac{\rho}{2}-
\frac{\delta^2}{\rho'}
\left(\frac{\omega}{q}+\frac12\rho\rho'\delta^2\right)
\frac{Q^2}{v_0}, \\
V_{T'}  &= \frac{1}{\rho'}
\left(1-\frac{\omega\rho'}{q}\delta^2\right)
\frac{Q^2}{v_0},
\end{align}
where we use the following definitions: $\delta = m_\mu/\sqrt{Q^2}$ (with $m_\mu$ being the muon mass), $\rho = Q^2/q^2$, and $\rho' = q/(\epsilon + \epsilon')$.

\begin{figure}[htp]
\centering
\includegraphics[scale=0.85, bb=6 204 544 663]{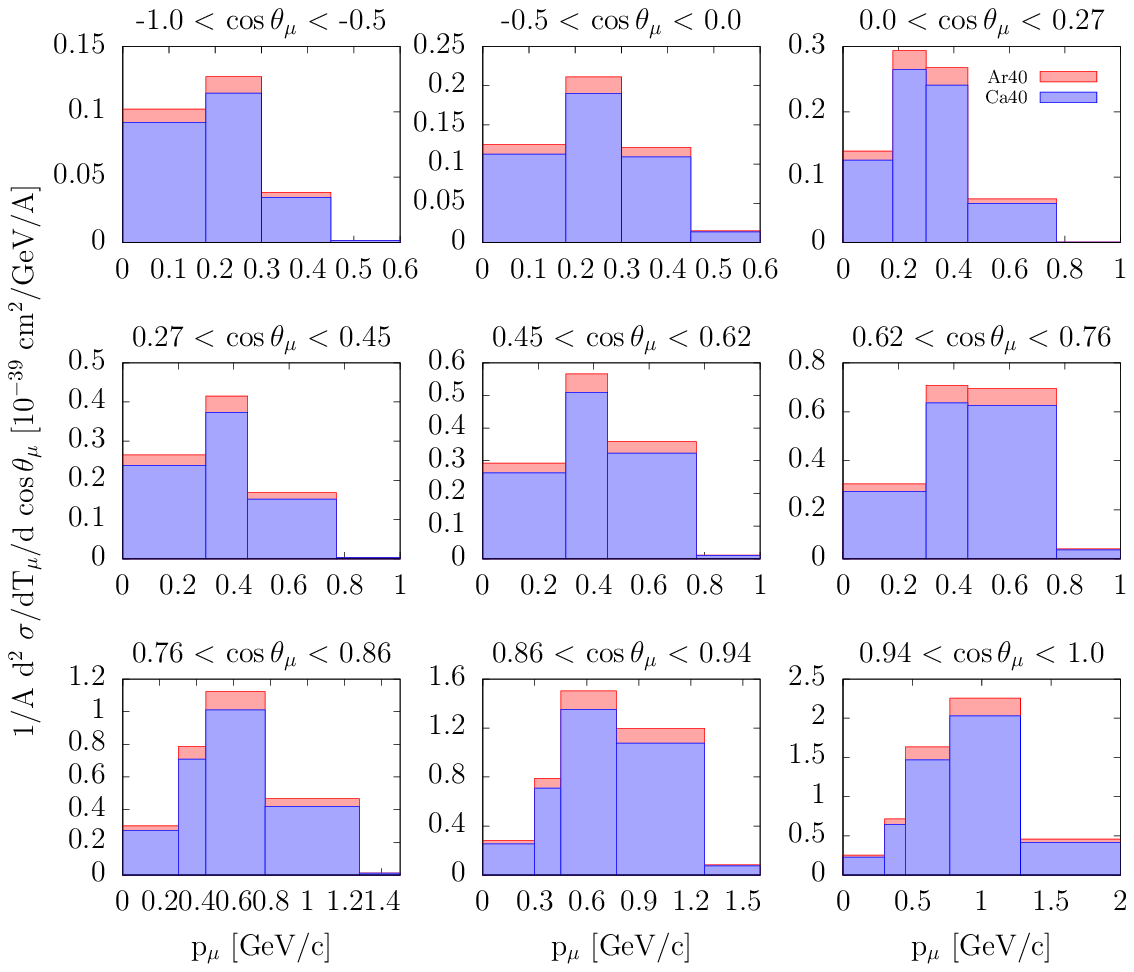}
\caption{MicroBooNE double-differential cross section for CC neutrino scattering on $^{40}$Ar and $^{40}$Ca, showing only the 2p2h-MEC contribution. This contribution is obtained by rescaling $^{12}$C results with the scaling formula developed in this work.}
\label{figura8}
\end{figure}

To connect with experimental observables, we focus on the $(\nu_\mu, \mu^-)$ channel on $^{40}$Ar under MicroBooNE-like kinematics. The measured quantity is the flux-averaged double-differential cross section, computed as
\begin{equation}
\left\langle \frac{d^2\sigma}{dT_\mu d\cos\theta_\mu} \right\rangle
=
\frac{
\int dE_\nu \, \Phi(E_\nu)
\frac{d^2\sigma
}{dT_\mu d\cos\theta_\mu}(E_\nu)}
{\int dE_\nu \, \Phi(E_\nu)},
\end{equation}
where $\Phi(E_\nu)$ represents the neutrino flux, and the numerator is the cross section at fixed neutrino energy.
A direct theoretical comparison of the 2p2h MEC responses in $^{40}$Ar and $^{40}$Ca is presented in Fig.~\ref{figura8}, under kinematic conditions representative of the MicroBooNE experiment. The analysis provides a detailed bin-by-bin study of the two nuclei, which share the same mass number $A=40$. A systematic deviation of about 10\% is observed between the $^{40}$Ar and $^{40}$Ca responses, in agreement with expectations and consistent with the trends discussed in previous sections.

\begin{figure}[htp]
\centering
\includegraphics[scale=0.85, bb=6 204 544 663]{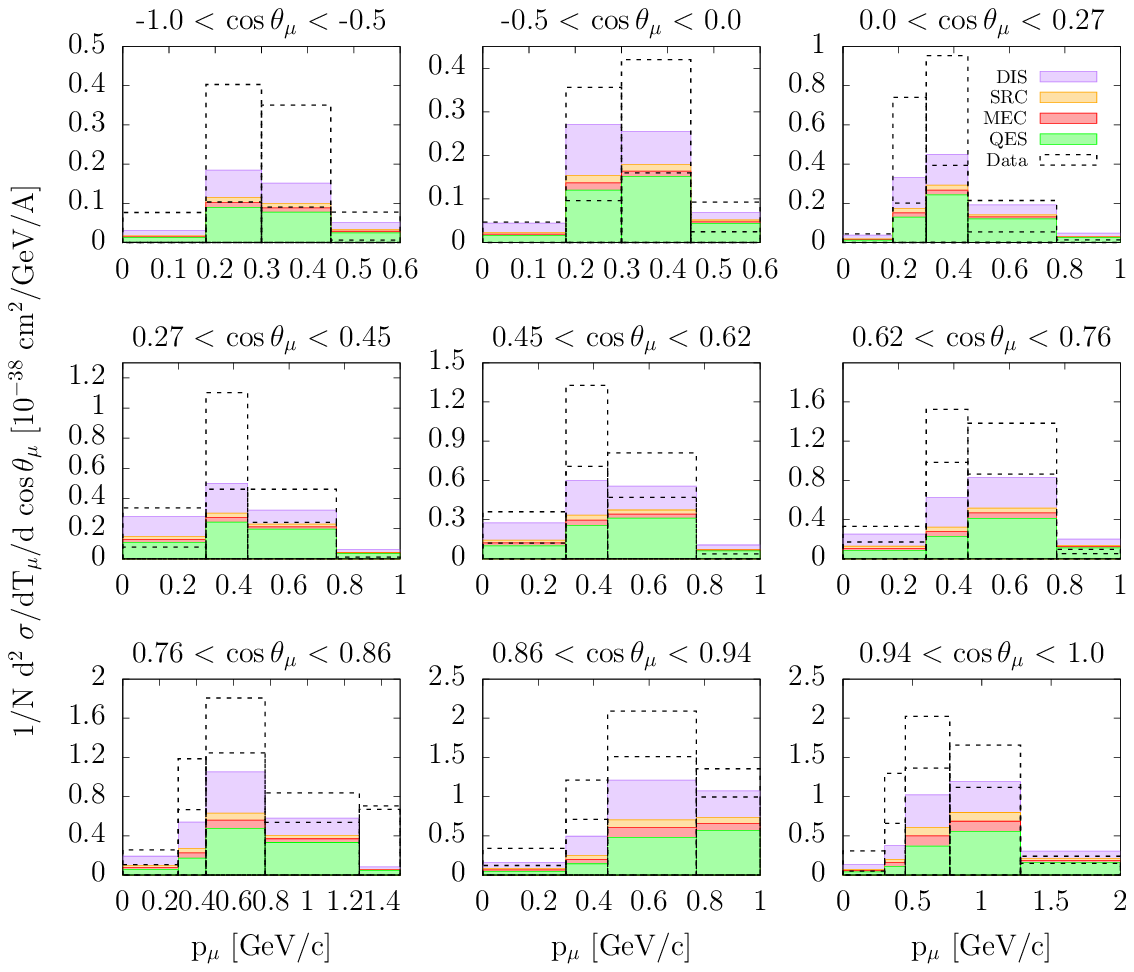}
\caption{Double differential cross section of CC neutrino scattering on an argon-40 target. Experimental data are taken from~\cite{Micro19}.}
\label{figura9}
\end{figure}

The flux-averaged charged-current neutrino cross section on $^{40}$Ar is shown in Fig.~\ref{figura9}, calculated under the same kinematic setup as in Fig.~\ref{figura8}. As in Fig.~\ref{figura7}, all reaction mechanisms are included---quasielastic, two-particle-two-hole meson-exchange currents, short-range correlations, and deep inelastic scattering ---providing the full cross section relevant for comparison with experiment. The experimental data are depicted as dashed rectangles, with the associated uncertainty represented by two horizontal dashed lines for each bin. To enable a meaningful comparison with the MicroBooNE measurements, the theoretical predictions have been folded with the experimental smearing matrix, which accounts for detector effects. This procedure allows a quantitative validation of the model against the observed data.

Overall, the theoretical model, taking the sum of all contributions, shows good agreement with the experimental data across most of the kinematic range. Deviations appear at small muon scattering angles ($\theta_\mu$) and low momentum transfer, where the model tends to underestimate the measured cross section. This discrepancy is primarily due to the DIS contribution, while the QE, MEC, and SRC contributions appear to be accurately described. The results shown in Fig.~\ref{figura9} are similar to the ones of Ref. ~\cite{jesus2}.

It is important to note that all 2p2h results shown in Figs.~\ref{figura7}, \ref{figura8}, and \ref{figura9} have been obtained by applying our asymmetric scaling formulas, constructing the 2p2h responses for $^{40}$Ar and $^{40}$Ca from the calculations of the $^{12}$C nucleus. We incorporate the proton-neutron asymmetry in all contributions included in the model, not just in the 2p2h responses.

\section{Conclusions}\label{section6}
We have presented a microscopic evaluation of two-particle-two-hole meson-exchange current responses in asymmetric nuclei, with emphasis on $^{40}\mathrm{Ar}$ and comparison to $^{40}\mathrm{Ca}$. By extending the relativistic mean-field and relativistic Fermi gas frameworks to treat protons and neutrons with distinct Fermi momenta, we have achieved a more realistic description than the commonly used symmetric matter approximation.

Quantitatively, we have found that the common practice of using $^{40}\mathrm{Ca}$ as a proxy for $^{40}\mathrm{Ar}$ induces an underestimation of about $10\%$ in the total 2p2h strength. The bias is channel-dependent: differences of around $6\%$ in the $pp$ channel can grow to $20$-$30\%$ in the $np$ channel, consistent with the behavior observed in the response functions. A density-based prescription reproduces RMF results reasonably well, but in the RFG framework (as used in SuSAv2+MEC) for Argon, deviations can become substantially larger (up to $\sim 38\%$), mainly because it does not explicitly incorporate proton-neutron asymmetry or separation-energy effects.

In contrast, the asymmetric scaling formulas presented in this work have captured the leading nuclear dependencies (asymmetry with $Z \neq N$, $k_{Fp} \neq k_{Fn}$) and reproduce $^{40}\mathrm{Ar}$ from $^{12}\mathrm{C}$. The main dependence is on $k_F^3$ (protons or neutrons), while residual nuclear dependencies are encoded in a scaling factor obtained from the microscopic calculation of $^{40}\mathrm{Ar}$ and $^{12}\mathrm{C}$. The asymmetric scaling formula assumes that the per-nucleon ratios remain approximately constant over the kinematic range of interest for the cross section. The coefficients have been tabulated for $pp$, $np$, and $nn$ pairs in both electron and neutrino scattering, and are the same for all nuclear responses.

We have presented predictions for both electron and neutrino cross sections on Argon. For direct applicability to measurements, the neutrino results have been folded with the detector smearing matrix provided by the MicroBooNE Collaboration, enabling comparisons at the reconstructed distribution.

The framework can be readily extended to other targets, including more neutron-rich and heavier nuclei. Its analytic formulation facilitates straightforward implementation in widely used neutrino event generators. These results may not directly translate to other models without performing the corresponding calculations within their respective frameworks to obtain the reduced coefficients.

Future work will focus on extending the asymmetric scaling formulas to a broader range of nuclei and on determining the dependence on proton and neutron Fermi momenta to obtain the dependence of the factors on the fermi momenta. The present scaling prescription can be directly applied in generators such as GENIE, NuWro, and NEUT, providing a practical tool to reduce systematic uncertainties in neutrino–nucleus interaction modeling, with particular relevance for experiments on argon targets, such as MicroBooNE and DUNE.
\begin{acknowledgments}
The authors would like to thank Guillermo Megias, Raul Gonzalez, and Jesus Gonzalez-Rosa for valuable discussions.

The work was supported by Grant No. PID2023-147072NB-I00 funded by MICIU/AEI /10.13039/501100011033 and by ERDF/EU; by Grant No. FQM-225 funded by  Junta de Andalucia;
V.L.M.C.  acknowledges financial support provided by  Ministerio Español de Ciencia, Innovación y Universidades under grant No. PID2022-140440NB-C22; Junta de Andalucía under contract Nos. PAIDI FQM-370 and PCI+D+i under the title: ``Tecnologías avanzadas para la exploración del universo y sus componentes'' (Code AST22-0001).
\end{acknowledgments}
\begingroup
    \renewcommand{\section}[2]{}

\endgroup

\end{document}